\documentclass[published]{epl}
\usepackage{epsfig}
\usepackage{amsmath,amssymb}
\usepackage{url}

\issue{58}{6}{2002}{1}{\parbox[t]{76.5mm}{\raggedleft arXiv version
    with permission from EDP Sciences\\\vspace{1.38ex}
    \url{http://www.edpsciences.org/euro/}}}

\makeatletter 
\renewcommand{\epl@received}{\relax}
\makeatother
\AtBeginDocument{%
  \makeatletter
  \newlabel{epl@first@page}{{}{799}}
  \newlabel{epl@last@page}{{}{805}}
  \makeatother
}
\newcommand{\e}[1]{{\mathbf e}_{#1}}
\newcommand{\x}{{\mathbf x}}
\newcommand{\y}{{\mathbf y}}
\newcommand{\z}{{\mathbf z}}

\newcommand{\yp}{{\mathbf y'}}
\newcommand{\zp}{{\mathbf z'}}
\newcommand{\E}{E}
\newcommand{\KS}{\E_{\text{KS}}}

\newcommand{\pfbox}{\hfill$\square$\qquad}

\title{A Kochen-Specker inequality}
 
\author{Jan-{\AA}ke Larsson\footnote{E-mail: \email{jalar@mai.liu.se}}} 

\institute{ Matematiska Institutionen, Link{\"o}pings Universitet,
  SE-581 83 Link{\"o}ping, Sweden}


\begin{document}

\maketitle

\begin{abstract}
  By probabilistic means, the concept of contextuality is extended so
  that it can be used in non-ideal situations.  An inequality is
  presented, which at least in principle enables a test to discard
  non-contextual hidden-variable models at low error rates, in the
  spirit of the Kochen-Specker theorem.  Assuming that the errors are
  independent, an explicit error bound of 1.42\% is derived, below
  which a Kochen-Specker contradiction occurs.
\end{abstract}

The description of quantum-mechanical processes by hidden variables is
a subject being actively researched at present. The interest can be
traced to topics where recent improvements in technology has made
testing and using quantum processes possible. Research in this field
is usually intended to provide insight into whether, how, and why
quantum processes are different from classical processes. Here, the
presentation will be restricted to the question whether there is a
possibility of describing a certain quantum system using a
\emph{non-contextual} hidden-variable model or not.  A non-contextual
hidden-variable model would be a model where the result of a specific
measurement does not depend on the context, i.e., what other
measurements that are simultaneously performed on the system.  It is
already known that in the ideal case no non-contextual model exists.
These results can be traced to the work of Gleason \cite{Gleason}, but
a conceptually simpler proofs was given by Kochen and Specker (KS)
\cite{KochSpeck}.

The original KS theorem concerns measurements on a quantum system
consisting of a spin-1 particle. In the quantum description of this
system, the operators associated with measurement of the spin
components along orthogonal directions do not commute, i.e.,
\begin{equation}
  \hat{S}_\x, \hat{S}_\y,\text{ and }\hat{S}_\z\text{ do not commute.}
\end{equation}
however, the operators that are associated with measurement of the
square of the spin components do commute, i.e.,
\begin{equation}
  \hat{S}^2_\x, \hat{S}^2_\y,\text{ and }\hat{S}^2_\z\text{ commute.}
\end{equation}
The latter operators (the squared ones) have the eigenvalues 0 and 1,
and 
\begin{equation}
  \hat{S}^2_\x+\hat{S}^2_\y+\hat{S}^2_\z=2\mathbb{I}.
\end{equation}
Thus, it is possible to simultaneously measure the square of the spin
components along three orthogonal vectors, and two of the results will
be 1 while the third will be 0. Only this quantum-mechanical property
of the system will be used in what follows. A hidden-variable model of
this system is now non-contextual if the result when measuring
$\hat{S}^2_\x$ is independent of whether $\hat{S}^2_\y$ and
$\hat{S}^2_\z$ are measured simultaneously or $\hat{S}^2_\yp$ and
$\hat{S}^2_\zp$ are. That is, measurement along one direction should
yield the same result irrespective of the other (commuting)
measurements made, so that the result of a measurement does not depend
on the context. The idea of the KS theorem is to choose a set of
orthonormal triads so that assignment of 0/1 results to the
corresponding measurements is impossible.

However, idealizations are needed in the KS theorem that are not
usable in experimental situations. The most obvious idealization is
that the measurement results are assumed to be without errors. Also,
it is assumed that all measurements yield results (no missing
detections). Further, a more subtle point is whether it is at all
possible to align two measurement setups so that there is one common
measurement, being made in different contexts (this question is raised
by Meyer, Kent and Clifton (MKC) \cite{MKC}). Lastly, it is often
stated that to detect changes due to the context, we need to prepare
several systems so that the hidden variables are identical. Here, a
probabilistic analysis of the theorem will be provided, addressing
these issues so that, at least in principle, an experimental test will
be possible. But let us first look at the ideal case.
 
To describe the hidden-variable model, standard probability theory
will be used, where the hidden variable $\lambda$ is a point in a
probabilistic space $\Lambda$, and sets in this space (``events'')
have a probability given by the probability measure $P$. The
measurement results are described by random variables (RVs)
$X_i(\lambda)$, which take their values in the value space
$V=\{0,1\}$. To describe the measurement results, RVs are used so that
formally the results are
\begin{equation}
  \begin{pmatrix}
    X_1(\x,\y,\z,\lambda)\\
    X_2(\x,\y,\z,\lambda)\\
    X_3(\x,\y,\z,\lambda)
  \end{pmatrix},\label{eq:lhv}
\end{equation}
which can take the results 0 or 1.  The RV $X_1$ is associated with
$\x$, $X_2$ with $\y$, and $X_3$ with $\z$; in a non-contextual
hidden-variable model, $X_1$ would not depend on the arguments $\y$
and $\z$, for example. To shorten the notation, the following
symmetries of the measurement results are assumed to hold (the proofs
go through without the symmetry, but grow notably in size):
\begin{equation}
  X_1(\x,\y,\z,\lambda)=
  X_2(\z,\x,\y,\lambda)=
  X_3(\y,\z,\x,\lambda).
\end{equation}
In the analysis, we will use a set of pairwise interconnected triads
 using the notation
\begin{equation}
  \E=\left\{
    \begin{pmatrix}
      \e1\\\e2\\\e3
    \end{pmatrix},
    \begin{pmatrix}
      \e1\\\e4\\\e5
    \end{pmatrix},
    \cdots,
    \begin{pmatrix}
      \e{n-2}\\\e{n-1}\\\e{n}
    \end{pmatrix}
  \right\}.
\end{equation}
In this set there are $n$ vectors forming $N$ distinct orthogonal
triads where some vectors are present in more than one triad,
establishing in total $M$ connections by rotation around a vector.
The KS theorem may now be stated.

\emph{Theorem 1: (Kochen-Specker)} There exists a set of vector triads
$\KS$, so that the following two prerequisites cannot hold
simultaneously for any $\lambda$
\begin{enumerate}
\item \emph{Non-contextuality.} For any pair of triads in $\KS$
  related by a rotation around a vector, the result along that vector
  is not changed by the rotation. For example,
  \begin{equation}
    X_1(\x,\y,\z,\lambda)=  X_1(\x,\y',\z',\lambda).
  \end{equation}
\item \emph{Quantum-mechanical results.} For any triad in $\KS$, the
  sum of the results is two, i.e.,
  \begin{equation}
    \sum_i X_i(\x,\y,\z,\lambda)=2.
  \end{equation}
\end{enumerate}

A proof of the Kochen-Specker (KS) theorem is essentially a
specification of $\KS$, and a check that (i) and (ii) indeed cannot
hold simultaneously. A full proof will not be provided here (the
original $\KS$ from \cite{KochSpeck} contains 117 vectors), but a
simplified walk-through is highly useful in what follows. The proof is
by contradiction; assume that Theorem~1~(i--ii) holds on $\KS$ for
some $\lambda$, and use (ii) to assign values to
$X_i(\e1,\e2,\e3,\lambda)$. Now, rotate using (i) to yield
$X_1(\e1,\e4,\e5,\lambda)$, and use (ii) to assign values to
$X_2(\e1,\e4,\e5,\lambda)$ and $X_3(\e1,\e4,\e5,\lambda)$. Rotate
again, and continue.  The value for many RVs at later triads in the
set will be determined by (i) and previous assignments, so there will
be less choices as we continue.  Now, when we arrive at the last triad
in $\KS$, all three values will already be fixed by (i) and previous
assignments so that
\begin{equation}
  \sum_i X_i(\e{n-2},\e{n-1},\e{n},\lambda)\neq 2,
\end{equation}
in contradiction to (ii). This will occur whatever choices we make in
our assignments (when we are free to make choices), which completes
the proof of the KS theorem.

Let us now generalize this theorem to take into account the different
kinds of non-ideal conditions occurring in experiment. The two first
problems presented above is easiest to manage. Errors can be of two
kinds: there can be (i) errors when changing context (e.g., a change
in $X_1$ when changing $(\y,\z)\to (\y',\z')$) or (ii) errors
in the sum. Missing results are also of two kinds: (i) when changing
context or (ii) in one or more of the three $X_i$s so that the sum
cannot be checked. These two can fairly easily be incorporated into a
generalized theorem, by estimating the rate of errors. The question
whether equal alignment in one direction is possible can be addressed
as follows: consider a measurement device as depicted in fig.~1,
presumably applied to a spin-1 system.

\begin{figure}[htbp]
  \begin{center}
    \psfig{file=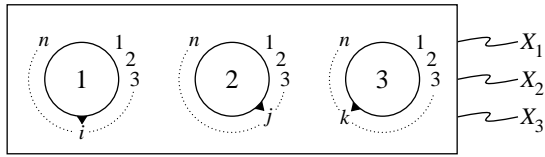}
    \caption{A Kochen-Specker experimental device. Inputs are the
      setting indices $(i,j,k)$, and outputs are $X_m$ taking the
      value 0 or 1, so that $\sum_m X_m=2$.}
  \end{center}
\end{figure}

In fig.~1, the (macroscopic, distinct) settings on the inputs indicate
which direction the measurements should be performed in. It is now
entirely possible that e.~g.\ the index-triple $(1,2,3)$ corresponds
to the triad $(\e1,\e2,\e3)^T$ while the index-triple $(1,4,5)$
corresponds to the triad $(\e1',\e4,\e5)^T$, so that \emph{the first
  vector changes although the first index does not}. This may be due
to a non-ideal measurement device, or perhaps to an MKC effect. The
effect on the results, however, is only a possible change in the
result $X_1$ even though the first \emph{index} has not changed, i.e.,
an error in (i) when changing the context. And errors of this type can
be incorporated as indicated above.

Lastly, it is often said that to show contextuality in an experiment,
it is necessary to prepare several systems so that the hidden
variables are identical; this is to make a point-wise analysis (in
$\Lambda$) of the condition (i).  The below theorem is a statistical
theorem that provides a \emph{global} analysis on the whole sample set
$\Lambda$, much in the same spirit as in \cite{IDENTICAL}, where this
issue also is examined. The statement of Theorem~2 may seem slightly
awkward; I have chosen to preserve the structure of Theorem~1 (note
the similarities of (i-ii) in Theorems~1 and~2). A simpler statement
will be provided in a later corollary.

\emph{Theorem 2: (Kochen-Specker inequality)} Given a KS set $\KS$ of
$N$ vector triads with $M$ interconnections by rotation, if
\begin{equation}
  M\delta+N\epsilon < 1,
\end{equation}
the following two bounds cannot hold simultaneously.
\begin{enumerate}
\item \emph{``Rotation'' error bound.} For any pair of triads in $\KS$
  related by a rotation around a vector, the set of $\lambda$s where
  the result along that vector is not changed by the rotation is large
  (has probability greater than $1-\delta$).  For example,
  \begin{equation}
    P\bigg(\Big\{\lambda:X_1(\x,\y,\z,\lambda)=
    X_1(\x,\y',\z',\lambda)\Big\}\bigg)\geq1-\delta.
  \end{equation}
\item \emph{``Sum'' error bound.}  For any triad in $\KS$, the set of
  $\lambda$s where the sum of the results is two is large (has
  probability greater than $1-\epsilon$), i.e.,
  \begin{equation}
    P\bigg(\Big\{\lambda:
    \sum_i X_i(\x,\y,\z,\lambda)= 2\Big\}\bigg)
    \geq 1-\epsilon.
  \end{equation}
\end{enumerate}

\emph{Proof:} 
We have (using $\complement$ to denote complement)
\begin{gather}
  \refstepcounter{equation}
  P\bigg(\Big\{\lambda:X_1(\x,\y,\z,\lambda)=
  X_1(\x,\y',\z',\lambda)\Big\}^{\complement}\bigg)\leq\delta,
  \tag{\theequation a}\\
  P\bigg(\Big\{\lambda:
  \sum_i X_i(\x,\y,\z,\lambda)= 2\Big\}^{\complement}\bigg)
  \leq \epsilon,\tag{\theequation b}
\end{gather}
so that,
\begin{align}
    &P\bigg(\Big(\underset{M}{\bigcap}\Big\{\lambda:
    X_1(\x,\y,\z,\lambda)=X_1(\x,\y',\z',\lambda)\Big\}\Big)
    \bigcap\Big(\underset{N}{\bigcap} \Big\{\lambda:
    \sum_i X_i(\x,\y,\z,\lambda)= 2\Big\} \Big)\bigg)\notag\\
    &=1-P\bigg(\Big(\underset{M}{\bigcup}\Big\{\lambda:
    X_1(\x,\y,\z,\lambda)=X_1(\x,\y',\z',\lambda)
    \Big\}^{\complement}\Big)
    \bigcup\Big(\underset{N}{\bigcup} \Big\{\lambda:
    \sum_i X_i(\x,\y,\z,\lambda)= 2\Big\}^{\complement}
    \Big)\bigg)\notag\\
    &\geq 1-\sum_M P\Big(\Big\{\lambda:
    X_1(\x,\y,\z,\lambda)=X_1(\x,\y',\z',\lambda)
    \Big\}^{\complement}\Big)
    -\sum_N P\Big( \Big\{\lambda:
    \sum_i X_i(\x,\y,\z,\lambda)= 2\Big\}^{\complement}\Big)\notag\\
    &\geq 1-M\delta-N\epsilon.\label{eq:PKS}
\end{align}
The intersection set in the first probability at the top is empty by
Theorem~1, i.e., the probability has to be zero. Thus, when the last
expression is strictly positive (when $M\delta+N\epsilon<1$), we have
a contradiction.  \pfbox

Note the explicit use of complement to take care of situations when
there is no measurement result. Thus, in Theorem~2, one need not
assume that the detected statistics correspond to that of the whole
ensemble (this is commonly known as the ``no-enhancement
assumption''). Unfortunately inefficient detector devices would
contribute no-detection events to both the error rates $\delta$ and
$\epsilon$, which puts a rather high demand on experimental
equipment. While the no-enhancement assumption can be used in
inefficient setups, this may weaken the statement (cf.\ a similar
argument for the GHZ paradox \cite{GHZ}). Also, as we shall see below,
even when using the no-enhancement assumption here, the demands on
experimental equipment are high.

The error rate $\epsilon$ is the probability of getting an error in
the sum (both non-detections and the wrong sum are errors here), not
the probability of getting an error in an individual result, and this
makes it easy to extract $\epsilon$ from experimental data.
Unfortunately, the errors that arise in rotation are not available in
the experimental data so it is not possible to estimate the size of
$\delta$ (note that it is not even meaningful to discuss $\delta$ in
quantum mechanics). It \emph{is} possible to use $\epsilon$ to obtain
a bound for $\delta$:

\emph{Corollary 3 (Kochen-Specker inequality)}
Given a KS set of $N$ vector triads $\KS$ with $M$ interconnections
by rotation, if Theorem~2 (i-ii) hold, then
\begin{equation}
  \delta\geq\frac{1-N\epsilon}{M}.
\end{equation}

As can be seen above, a small $\KS$ set (small $N$ and $M$) is better,
yielding a higher bound for $\delta$ for a given $\epsilon$. In an
inexact experiment yielding a large $\epsilon$, the bound in
Corollary~3 will be low. Being an inexact experiment, one expects the
error rate $\delta$ to be large also, thus not violating Corollary~3;
changes in a result due to the context may be attributed to (normal)
measurement errors. A model for this inexact experiment can then be
said to be ``probabilistically non-contextual''. In a better
experiment yielding a low $\epsilon$, the bound in Corollary~3 is
higher. But here one expects $\delta$ to be low as well, in violation
of Corollary~3. In a hidden-variable model of this experiment, the
changes in a result due to the context occur at an unexpectedly high
rate which cannot be attributed to measurement errors, and a model of
this type can be said to be ``probabilistically contextual''.

A small further discussion of the MKC \cite{MKC} constructions is
appropriate here (see also \cite{ONMKC}). The argument is that due to
the finite precision of measurement device orientations, we may only
get measurement results from a (dense) subset of vectors on the unit
sphere.  One of the constructions then proceeds to devise a dense
subset of vectors on the unit sphere such that Theorem~1 cannot be
applied. Results may now be assigned to each measurement following the
quantum rule (ii) on the whole dense subset, without any contradiction
(the other constructions are similar). Now, the MKC models are argued
to be non-contextual in the sense that only the vector along which the
measurement is made is needed to fix the value of the RV. An important
observation is that in the MKC models there is only one context in
which each measurement is contained; mathematically in (i), the set
$\{\lambda:X_1(\x,\y,\z,\lambda)=X_1(\x,\y',\z',\lambda)\}$ is
\emph{not well-defined}; there are no distinct triads $(\x,\y,\z)$,
$(\x,\y',\z')$ that can be inserted as arguments. Therefore, it can be
argued that the question of (non-)contextuality is void; an MKC model
is \emph{neither} contextual nor non-contextual. Also note that
Theorem~2 applies to any measurement device like that in fig.~1. Thus,
even if the discussion can be continued as to whether contextuality is
a valid concept for the MKC models at an internal level, one is forced
to conclude that a hidden-variable model (including the measurement
device) of a ``good'' experiment must be (probabilistically)
contextual.

Until now, the present discussion has been limited to the original
spin-1, 3-dimensional setting. However, there are higher-dimensional
settings in which there are smaller $\KS$, and Theorem~2 and
Corollary~3 hold for these as well. All that is needed is a simple
restatement of the prerequisites, i.e., in $d$ dimensions, the set
$\E$ consists of $N$ $d$-tuples with $M$ interconnections, and (i--ii)
need additional vector entries as in
$X_i(\e{1},\ldots,\e{d},\lambda)$. Finally, in (ii) a more general
specification of the results is that there should be one ``zero'' and
the rest ``ones'', so that the sum is $d-1$ (note that the labeling of
the results differ in the literature).

Returning to the proof of Theorem~2, to obtain a general statement,
there are no assumptions on independence of the errors, but it is
possible to give a more quantitative bound for the error rate by
introducing independence (for simplicity, using the no-enhancement
assumption). Please note that there is no experimental check whether
the assumption of independent errors holds or not. While the errors in
the sum may be possible to check, it is not possible to extract what
errors are present in the rotations or check for independence of those
errors \cite{INDEP}. The below result is dimension-dependent; $d$ is
the dimension.

\emph{Corollary 4 (KS inequality for independent errors):} Assuming
that the errors are independent at the rate $r$, both $\delta$ and
$\epsilon$ are given by $r$. If Theorem~2 (i-ii) hold, then
\begin{equation}
  M(2r-2r^2)+N(1-(1-r)^d-(d-1)(1-r)^{d-2}r^2) \geq 1.
\end{equation}

\emph{Proof:} In the case of independent errors at the rate $r$, the
expressions for the probabilities in Theorem~2~(i) and (ii) are
\begin{gather}
  \refstepcounter{equation}
    \begin{split}
      P\bigg(\Big\{\lambda:
      X_1(\x,\y,\z,\lambda)=X_1(\x,\y',\z',\lambda)\Big\}\bigg)
      &=P(\text{no errors})+P(\text{flip on both }X_1\text{'s})\\
      =(1-r)^2+r^2
      &=1-(2r-2r^2),
    \end{split}\tag{\theequation a}\\
    \begin{split}
      P\bigg(\Big\{\lambda:
      \sum_i X_i(\x,\y,\z,\lambda)= 2\Big\}\bigg)
      &=P(\text{no errors})+P(\text{flip of the 0 and one 1})\\
      =(1-r)^d+(d-1)(1-r)^{d-2}r^2
      &=1-(1-(1-r)^d+(d-1)(1-r)^{d-2}r^2).
    \end{split}\tag{\theequation b}
\end{gather}
The probabilities of these sets are \emph{not} independent, so from
this point on we cannot use independence, but the result follows
easily from Theorem~2. \pfbox

An expression of the form $r\geq f(N,M,d)$ can now be derived from
Corollary~4, but this complicated expression is not central to the
present paper.  Again, to obtain a contradiction for high error rates
$r$, a small $\KS$ set is needed (small $N$ and $M$). It is now only a
matter of looking at the various $\KS$ available in the literature to
determine $N$ and $M$, however, some care should be taken when doing
this. 

For example, the set presented in \cite{Peres} contains 33 vectors
forming 16 distinct triads. Unfortunately, some rotations used in the
proof chain are not between two of these 16 triads; in some cases a
rotation goes from one of the 16 triads to a pair of vectors (where
the third vector needed to form a triad is not in the set), and a
subsequent rotation returns us to another of the 16 triads. On these
pairs, a modified result rule is used; there cannot be more than one
``zero'' in such a pair (see \cite{Peres} Section~7-3 or \cite{Bub}
Section~3.2). Translating to the notation used here, an extra vector
needs to be added in this case, and in the proof-chain in \cite{Peres}
Table~7-1, 8 extra vectors are needed; for symmetry reasons the total
number of added vectors is 24, yielding $n=57$ and $N=40$ (note that
these additional vectors are \emph{not} needed to yield the KS
contradiction if the modified result rule is used). Some care is also
needed to determine $M$ from the extended set, and for the set in
\cite{Peres} this yields a total of $M=96$ rotations.

To conclude, for any hidden-variable model we have a bound on the
rate $\delta$ of changes arising in rotation:
\begin{equation}
  \delta\geq\frac{1-N\epsilon}{M}.
\end{equation}
Here, $N$ is the number of triads in $\KS$, $M$ is the number of
connections within $\KS$, and $\epsilon$ is the error rate in the sum.
A proof using few triads with few connections yields a more
restrictive bound. An inexact measurement producing a large $\epsilon$
leads one to expect a large $\delta$, which is in accordance with the
inequality. A better experiment yielding a small $\epsilon$ leads one
to expect a small $\delta$, which is not possible; any hidden-variable
description of that physical system will have to be probabilistically
contextual (i.e., yield a larger rate of changes in rotation than one
would expect).

The above reasoning is qualitative, but if the assumption of
independent errors is used, an explicit bound can be determined for
the error rate $r$, and in the below table are a few examples (the $n$
column sometimes contains two numbers; the number in the parentheses
pertains to the original set and the other to the extended set
\cite{EXTEND}).

\begin{center}
  \begin{tabular}{l r c r r c}
    & $d$ & $n$ & $N$ & $M$ & $r$\\
    \hline
    Peres \cite{Peres} & 3 & 57 (33) & 40 & 96 & 0.0032\\
    Kochen \& Conway \cite{Peres} & 3 & 51 (31) & 37 & 91 & 0.0034\\
    Sch{\"u}tte \cite{Bub} & 3 & 49 (33) & 36 & 87 & 0.0035\\
    Kernaghan \& Peres \cite{KernaghanPeres} & 8 & 36 & 11 & 72 & 0.0043\\
    Kernaghan \cite{Kernaghan} & 4 & 20 & 11 & 30 & 0.0097\\
    Cabello \emph{et al} \cite{CGA} & 4 & 18 & 9 & 18 & 0.0142
  \end{tabular}
\end{center}

The $r$ column contains the highest value for which a contradiction to
(probabilistic) contextuality occurs. Obviously, a small KS set is
better than a large one, yielding a contradiction at a higher $r$.
Somewhat unexpectedly, Sch{\"u}tte's 33-vector set is more economical in
this sense than Kochen and Conway's 31-vector set. In the above table
we can see that there is a contradiction for the $\KS$ from \cite{CGA}
as soon as $r<1.42\%$. 

While writing this paper, the author learned from C.~Simon that a
similar approach was in preparation by him, \v{C}.\ Brukner, and A.\
Zeilinger \cite{CBZ}. It should be noted that in that paper, $\delta$
is implicitly assumed to be zero.
\stars

The author would like to thank A.\ Kent for discussions. This work was
partially supported by the QIT Programme in the European Science
Foundation.


\begin{thebibliography}{1}
\bibitem{INDEP} A discussion of the assumption of independent errors
  is appropriate here, but such a discussion cannot be fit into this
  limited space.
\bibitem{EXTEND} It is possible in the cases when an extended set is
  needed, that an analysis more adapted to the structure of the
  original set yields a contradiction at a somewhat higher $r$.
  Please also note that the observation by Clifton reported in
  \cite{Bub,CGA} that some vectors can be left out in
  \cite{Kernaghan,CGA} is similar to pruning the extra vectors in the
  extension of e.g.\ Peres's set \cite{Peres}.
\end{thebibliography}
\end{document}